\documentclass{aa}  

\usepackage{graphicx}
\usepackage{txfonts}
\usepackage{orcidlink}

\usepackage{amssymb}

\usepackage{placeins}

\begin{document}

   \title{The Impact of Magnitude Uncertainties and K-corrections on Standard Siren Measurements of the Hubble Constant}
   \titlerunning{The Impact of Magnitude Uncertainties and K-corrections on Standard Siren Measurements of the $H_0$}

   %\subtitle{I. Overviewing the $\kappa$-mechanism}

   \author{M. Pálfi \orcidlink{0000-0001-5942-0470} \thanks{e-mail: marika97@student.elte.hu}
         \inst{1}
          \and
          P. Raffai\orcidlink{0000-0001-7576-0141}\inst{1,2}\fnmsep
          }
    \authorrunning{M. Pálfi and P. Raffai}

   \institute{Institute of Physics and Astronomy, ELTE E\"otv\"os Lor\'and University, 1117 Budapest, Hungary
         \and
             HUN-REN–ELTE Extragalactic Astrophysics Research Group, 1117 Budapest, Hungary
            }

   %\date{Received XXX; accepted YYY}
   \date{}

% \abstract{}{}{}{}{} 
% 5 {} token are mandatory
 
\abstract
   {Gravitational-wave standard sirens provide an independent probe of the Hubble constant. Dark siren analyses that rely on galaxy catalogues are sensitive to catalogue-related uncertainties. While incompleteness, weighting schemes, and redshift errors have been studied extensively, the impact of magnitude uncertainties and K-corrections has received little attention so far.}
   {We assess how  magnitude errors and K-corrections propagate into the Hubble constant inference, and compare their impact to that of realistic spectroscopic redshift errors to determine their relevance.}
   {We use a modified version of the \texttt{gwcosmo} Python package to construct line-of-sight redshift priors from a complete, volume-limited mock galaxy catalogue up to $z<0.2$. We simulate and analyse a sample of $\sim200$ binary black hole events with luminosity weighting in the $B$- and $K$-bands and test different uncertainty models for redshifts, magnitudes, and K-corrections.}
   {The choice of luminosity-weighting scheme has a substantial impact on the inferred Hubble constant posterior, with the direct $B$- and $K$-band comparison producing differences comparable to the largest uncertainty-induced changes. Spectroscopic redshift errors produce the largest integrated changes in the Hubble constant posterior among the tested uncertainty treatments. Photometric magnitude uncertainties lead to substantially smaller, band-dependent deviations. In the $K$-band, the tested K-correction treatments have only a minor impact in the present low-redshift mock analysis, with the empirical correction slightly reducing the integrated deviation relative to the no-correction case.}
   {Our results show that, for a complete, low-redshift mock catalogue, magnitude uncertainties and K-correction assumptions are subdominant compared to spectroscopic redshift errors, while the luminosity-weighting scheme represents an important modelling choice in catalogue-based dark siren inference. In realistic, flux-limited catalogues with larger localization areas and photometric or mixed redshift errors, the relative impact of magnitude uncertainties is expected to be even smaller, indicating that they need not be prioritized in near-future dark siren analyses.}

   \keywords{Gravitational waves -- Cosmology: cosmological parameters
                 -- Methods: numerical
               }

   \maketitle
   \nolinenumbers

%
%________________________________________________________________

\section{Introduction}

The detection of GW170817 \citep{GW170817} and its electromagnetic counterpart \citep{GW170817_EM} marked the beginning of a new era in astronomy, the era of multimessenger astronomy with gravitational waves. This was the first event for which both the luminosity distance from the gravitational-wave (GW) signal and the redshift from the identification of the host galaxy became available, enabling the first bright siren measurement of the Hubble constant, $H_0$ \citep{GW170817_BS}. For dark sirens -- compact binary coalescences  without electromagnetic counterparts -- the host galaxies cannot be identified, and therefore statistical methods are needed to obtain redshift information \citep{Schutz1986, cosmo_review}. Several analysis frameworks have been developed for dark siren cosmology, including \texttt{icarogw} \citep{Mastrogiovanni_2024, Mastrogiovanni_2023} and \texttt{CHIMERA} \citep{Borghi_2024, Tagliazucchi_2025}. One widely used solution is to utilize galaxy catalogues to statistically associate the GW event with potential host galaxies and their redshifts \citep{DS_1, DS_2, Hitchhiker}. This galaxy-catalogue-based dark siren method was already implemented in early works such as \texttt{DarkSirensStat} \citep{Finke_2021}.  It is also implemented in the publicly available \texttt{gwcosmo} software package \citep{gwcosmo3, gwcosmo2, gwcosmo1}, which provides a scalable Bayesian analysis framework for both bright and dark siren GW cosmology, allowing for the joint estimation of cosmological parameters and compact-binary population properties.

The posterior on $H_0$ can be biased or broadened by uncertainties arising both from the GW observations themselves \citep[e.g.,][]{mass_models, waveform} and from the galaxy catalogue used in the analysis. The impact of galaxy catalogue uncertainties and deficiencies on the estimation of the $H_0$ was first systematically explored by \cite{gwcosmo3}, based on a fully simulated universe with mock GW detections and synthetic galaxy data, analysed using the \texttt{gwcosmo} framework. The authors investigated the impact of catalogue incompleteness, luminosity-based galaxy weighting, and the number of detected events on the $H_0$ inference. They demonstrated that missing galaxies bias the results, proper luminosity weighting improves precision, and larger samples of GW events reduce statistical uncertainty. 

Weighting galaxies by their intrinsic likelihood of hosting a GW source can improve cosmological inference, as some galaxy populations are expected to be more likely hosts than others based on their stellar mass, star formation rate, or evolutionary history \citep[e.g.,][]{HGP,  Artale_1, Artale_2, Santoliquido, Inferring}. The effect of different luminosity weighting schemes on the $H_0$ was studied in \cite{GW170817_weights} using GW170817 as a dark siren, showing that such weighting can tighten constraints but introduces bias if misapplied. Using simulated data, \cite{Perna} showed that applying an incorrect host galaxy probability model -- one that ignores the dependence on galaxy luminosity and redshift -- can lead to biased results, highlighting the importance of adopting a physically motivated weighting scheme. Similarly, \cite{SM_SFR} demonstrated with mock galaxy catalogues that weighting galaxies by stellar mass or star formation rate can significantly bias the $H_0$ posterior if the weighting model does not match the true merger rate distribution. This result is reinforced by \citet{Alfradique_2025}.  Relatedly, \citet{UtilizingStellarMass} explored how stellar-mass-based weighting influences the identification of the true host galaxy using mock galaxy catalogues, providing further support for the use of physically motivated weighting schemes.

Peculiar velocities introduce an additional source of uncertainty, particularly at low redshifts where their relative contribution to the observed redshift is larger. In the case of GW170817, \cite{pecvel_GW170817} studied the impact of peculiar velocities and showed that modelling these correctly is essential to obtain an accurate $H_0$ estimate. \cite{pecvel} demonstrated that peculiar velocities correlate with stellar mass, evolve with redshift, and can bias the $H_0$ posterior if not properly accounted for, especially for nearby events. Uncertainties in galaxy redshift measurements also affect the estimation of the $H_0$, especially when using photometric redshifts or low-resolution spectroscopy, where the redshift errors are typically larger. \cite{redshifterr} studied the impact of redshift uncertainties using real data and various error models, which they implemented in the \texttt{gwcosmo} software package, showing that applying incorrect or overly simplistic or unrealistic error models can bias the $H_0$ inference. Using mock galaxy catalogues and assuming Gaussian luminosity-distance likelihoods, \citet{cross-parkin2025} investigated how redshift precision affects dark-siren cosmology, showing that spectroscopic-level redshift accuracy significantly improves $H_0$ constraints (particularly for well-localized events), while realistic redshift outliers — defined as cases where the estimated redshift strongly deviates from the true value — do not necessarily bias the inference. Using simulated O5-like GW events, mock galaxy catalogues, and the \texttt{CHIMERA} framework, \citet{Borghi_2024} further quantified the impact of redshift uncertainties, demonstrating that increasing redshift errors degrade the precision of the $H_0$ measurement and can introduce biases if not properly accounted for in the analysis.

Catalogue completeness is another key factor, as missing galaxies can bias the analysis if the true host lies outside the catalogue. \citet{Alfradique_2025} demonstrated with mock galaxies and simulated GW signals that ignoring catalogue incompleteness systematically shifts the $H_0$ estimate towards lower values. They further showed that different apparent-magnitude cuts affect the shape of the $H_0$ posterior, whereas using a volume-limited catalogue restores an unbiased result. \citet{completeness_test} introduced a statistical method to estimate the magnitude limit from the magnitude–redshift distribution of a galaxy catalogue and applied it in dark siren analyses, demonstrating that including fainter galaxies beyond the limit used originally in the \texttt{gwcosmo} method improved the $H_0$ inference when applied to real data. In contrast, \cite{bright_completeness} argued that restricting the analysis to the brightest, most complete part of the catalogue is an effective strategy for nearby events, where the brightest galaxies dominate and completeness is easier to achieve. More recently, \citet{Borghi_2026} jointly investigated the effects of galaxy catalogue incompleteness and host-galaxy weighting schemes using simulated GW events, mock galaxy catalogues, and the \texttt{CHIMERA} framework, showing that incorrect weighting has little impact for complete catalogues, but can introduce significant biases when the catalogue is incomplete.

While several studies have explored the impact of catalogue incompleteness, weighting schemes, peculiar velocities, and redshift errors, the effects of magnitude measurement uncertainties and K-corrections have received little attention in the context of dark siren cosmology. In particular, \cite{Abbott_2021} pointed out that neglecting K-corrections can lead to redshift-dependent systematics, while \cite{gwcosmo1} noted in a footnote that magnitude errors are usually neglected as being small, leaving their impact to future investigation. The general assumption,  also adopted in the publicly available \texttt{gwcosmo} package, is that the magnitude uncertainties are perfectly known. \cite{DS_GW170817} and \cite{Palmese_2020} included a fixed 0.05 mag uncertainty in their analysis where they treated GW170817 as a dark siren, but they did not investigate the effect of magnitude errors on the $H_0$ estimate. In this paper, we fill this gap by systematically studying how magnitude measurement errors and K-correction systematics propagate into the inference of the Hubble constant, and by assessing their relevance for upcoming analyses. To this end, we perform an end-to-end simulated analysis based on mock galaxy catalogues and simulated gravitational-wave events, using the publicly available \texttt{gwcosmo} framework.

The paper is organized as follows. In Section$~$\ref{sec:methods}, we describe our approach to modelling magnitude uncertainties and K-corrections, and outline the methodology used to assess their impact on the $H_0$ inference. We present the results of our analysis in Section$~$\ref{sec:results}, and finally, we summarize our findings and discuss their implications for future standard siren analyses in Section$~$\ref{sec:conclusions}.

\section{Methods}\label{sec:methods}

To assess the impact of magnitude uncertainties on cosmological inference, we adopt the hierarchical Bayesian framework introduced in \citet{gwcosmo1}, based on the formalism derived in \citet{Mandel19} and \citet{Vitale2020}, and implemented in the \texttt{gwcosmo} package. In this formalism, the posterior on hyperparameters $\Lambda$ (which include both cosmological and compact binary population parameters) is given by
\begin{align}
    p(\Lambda \mid \{x_{\mathrm{GW}}\}, \{D_{\mathrm{GW}}\}, & I) \propto \ p(\Lambda \mid I)\, p(N_{\mathrm{det}} \mid \Lambda, I) \nonumber \\
    &\times \prod_{i=1}^{N_{\mathrm{det}}} \frac{\int p(x_{\mathrm{GW}_i} \mid \theta, \Lambda, I)\, p(\theta \mid \Lambda, I)\, \mathrm{d}\theta}{\int p(D_{\mathrm{GW}_i} \mid \theta, \Lambda, I)\, p(\theta \mid \Lambda, I)\, \mathrm{d}\theta},
\end{align}
given $N_{\mathrm{det}}$ GW detections $\{D_{\mathrm{GW}}\}$ with corresponding data $\{x_{\mathrm{GW}}\}$. Here, $i$ indexes the individual events, $\theta$ denotes the intrinsic source parameters (including redshift $z$), and $I$ summarizes any additional assumptions. The denominator accounts for the detection probability of a compact binary coalescence drawn from the population described by $\Lambda$.

For dark sirens, the redshift distribution of potential host galaxies is incorporated via the so-called line-of-sight (LOS) redshift prior. This prior, marginalized over the galaxies’ absolute $M$ and apparent $m$ magnitudes (in the chosen photometric band), takes the form
\begin{align}\label{eq:catparts}
    p(z \mid \Omega_j, \Lambda, s, I) &=  \sum_{g = G,\,\overline{G}} p(g \mid \Omega_j, \Lambda, s, I) \nonumber \\
    & \quad \times \int\!\!\int p(z, m, M \mid g, \Omega_j, \Lambda, s, I)\, \mathrm{d}M\, \mathrm{d}m,
\end{align}
where $g = G$ refers to the case where the host galaxy is in the catalogue, and $g = \overline{G}$ corresponds to galaxies not included. The variable $s$ indicates that the GW signal is real (previously included in the assumptions $I$), and $\Omega_j$ denotes the $j$th sky pixel where the LOS $z$ prior is evaluated.

In the original implementation \citep{gwcosmo1}, the apparent magnitudes were assumed to be known exactly, modelled using $\delta$-functions. Here, we instead replace these with Gaussian distributions. When integrating the in-catalogue part over $M$, we have
\begin{align}
\int p(z, M, m \mid & G, \Omega_i, \Lambda, s, I)\, \mathrm{d}M \nonumber \\ &= \frac{p(z, m \mid G, \Omega_i, I)\, p(s \mid z, M(z, m, \Lambda), \Lambda, I)}{p(s \mid G, \Omega_i, \Lambda, I)},
\end{align}
with
\begin{align}
    p(z, m \mid G, \Omega_i, I) & =  \frac{1}{N_\text{gal}(\Omega_i)}  \nonumber \\ & 
    \quad \times \sum_{k=1}^{N_\text{gal}(\Omega_i)} p(z \mid \hat{z}_k, \hat{\sigma}_{z,k})\, p(m \mid \hat{m}_k, \hat{\sigma}_{m,k}),
\end{align}
where we assume independent Gaussian posteriors for the redshift $z$ and apparent magnitude $m$ of galaxy $k$,
\begin{align}
    & p(z \mid \hat{z}_k, \hat{\sigma}_{z,k}) = \mathcal{G}(z - \hat{z}_k, \hat{\sigma}_{z,k}), \\
    & p(m \mid \hat{m}_k, \hat{\sigma}_{m,k}) = \mathcal{G}(m - \hat{m}_k, \hat{\sigma}_{m,k})
\end{align}
with observed values $\hat{z}_k$ and $\hat{m}_k$, and $\hat{\sigma}_{z,k}$ and $\hat{\sigma}_{m,k}$ are the measurement uncertainties, modelled as the standard deviations of  the corresponding Gaussian distributions. Photometric data processing methods generally model magnitude uncertainties as Gaussian, based on the assumption of Gaussian-distributed flux measurements—see, e.g., SDSS \citep{Ivezic03}, 2MASS \citep{2MASS}, Pan-STARRS \citep{Pan-STARRS}, the Dark Energy Survey \citep{DES}, and LSST \citep{LSST}. In this study, we neglect potential correlations between redshift and magnitude uncertainties, which would otherwise require a joint posterior $p(z, m \,| \, \hat{z}_k, \hat{m}_k, \hat{\sigma}_{z,k}, \hat{\sigma}_{m,k})$. Substituting the above, the in-catalogue contribution becomes
\begin{align}
p(z|G,\Omega_i,\Lambda,&s,I)
= \int \!\! \int p(z,M,m|G,\Omega_i,\Lambda,s,I)\,dM\,dm \nonumber \\
&\quad \quad = \frac{1}{p(s|G,\Omega_i,\Lambda,I)N_\text{gal}(\Omega_i)}
\sum_{k=1}^{N_\text{gal}(\Omega_i)}
\Bigg[p(z|\hat{z}_k,\hat{\sigma}_{z,k}) \nonumber \\
&\times 
\int
p(m|\hat{m}_k,\hat{\sigma}_{m,k})
p(s|z,M(z,m,\Lambda),\Lambda,I)\,dm\Bigg] .
\label{eq:LOS}
\end{align}
The remaining integral over $m$ is evaluated numerically during the LOS-prior construction, thus the quantity  remains a one-dimensional redshift prior. The out-of-catalogue term remains unchanged, as it does not depend on measured galaxy properties (see \citealt{gwcosmo1}).

To generate a realistic dataset of GW sources and their host galaxies, we use the same mock catalogue and injection setup as in our previous work \citep{UtilizingStellarMass}. The galaxy catalogue was created with the Theoretical Astrophysical Observatory\footnote{\url{https://tao.asvo.org.au/tao/}} (TAO) using the Semi-Analytic Galaxy Evolution (SAGE) model \citep{Croton_2016}, based on light-cone outputs from the Millennium Simulation \citep{Springel2005}. The catalogue covers the full sky up to redshift $z = 0.2$. We computed absolute and apparent magnitudes for the 2MASS $K_s$ and WISE W1 and W2, and Johnson $B$-bands using the SED module of TAO, with the \citet{BC03} stellar population synthesis model and a \citet{Chabrier_2003} initial mass function. No dust attenuation or intergalactic absorption was applied. To balance completeness and computational feasibility, we selected galaxies with stellar masses above $10^8~\mathrm{M}_\odot/h$, resulting in approximately 78 million galaxies.

For the mock event generation, we drew 1000 host galaxies with sampling probability proportional to stellar mass, and we applied a redshift-dependent merger-rate weight following equation$~$(4) of \cite{O3cosmo}, based on the \cite{MandD} parametrization (using $\gamma=4.56$, $k=2.86$, $z_p=2.47$). Although host galaxies are sampled with probabilities proportional to stellar mass and a mild redshift-dependent merger-rate evolution (Madau–Dickinson), in our $z~<~0.2$ mock galaxy catalogue the latter varies little, thus stellar-mass weighting dominates the host sampling.  We simulated one binary black hole (BBH) merger per host and retained  them as GW detections if their network signal-to-noise ratio in an O4-like three-detector network, consisting of two Advanced LIGO detectors and Virgo, exceeded the threshold $\rho_\mathrm{net}=11$. This SNR threshold was the only GW detectability criterion and resulted in 195 detectable events, all of which were used in the cosmological inference.

We inferred the source parameters of these detected events using the \textsc{bilby} Bayesian inference library \citep{bilby_paper} with the \textsc{dynesty} nested sampler \citep{dynesty}, automated via the \textsc{bilby\_pipe} tool \citep{bilby_pipe_paper}. The BBH parameters were drawn from the same priors used in the inference (following the default BBH priors in \textsc{bilby}), except that the BBHs were placed in galaxies from the mock catalogue, i.e. their sky positions and luminosity distances matched those of the host galaxies. However, the parameter estimation was carried out using general astrophysical priors, without assuming knowledge of the true host. The GW signals were analysed with the precessing \texttt{IMRPhenomPv2} waveform approximant, using \texttt{lal\_binary\_black\_hole} as the frequency-domain source model. The posterior samples were then used to construct three-dimensional GW localization maps with \texttt{ligo.skymap}.

The GW selection-effect correction entering the dark-siren likelihood was calculated from the standard \texttt{gwcosmo} injection set, generated independently of the mock galaxy catalogue and selected according to the detector response and the same network-SNR threshold used to define the analysed mock-detection sample. The same default \texttt{IMRPhenomPv2} approximant was used for the \texttt{gwcosmo} injection set. In the subsequent \texttt{gwcosmo} inference and selection-effect correction, the BBH population was modelled with the \texttt{BBH-powerlaw-gaussian} mass model, i.e.~a power-law primary-mass distribution with an additional Gaussian peak and a power-law mass-ratio distribution.\footnote{
The fixed hyperparameters were 
$m_\mathrm{min}^{\rm BH}=4.98\,M_\odot$, 
$m_\mathrm{max}^{\rm BH}=112.5\,M_\odot$, 
$\alpha=3.78$, 
$\mu_g=32.27\,M_\odot$, 
$\sigma_g=3.88\,M_\odot$, 
$\lambda_g=0.03$, 
$\beta=0.81$, and 
$\delta_m=4.8\,M_\odot$.}

\begin{figure*}
    \centering
    \includegraphics[width=1\linewidth]{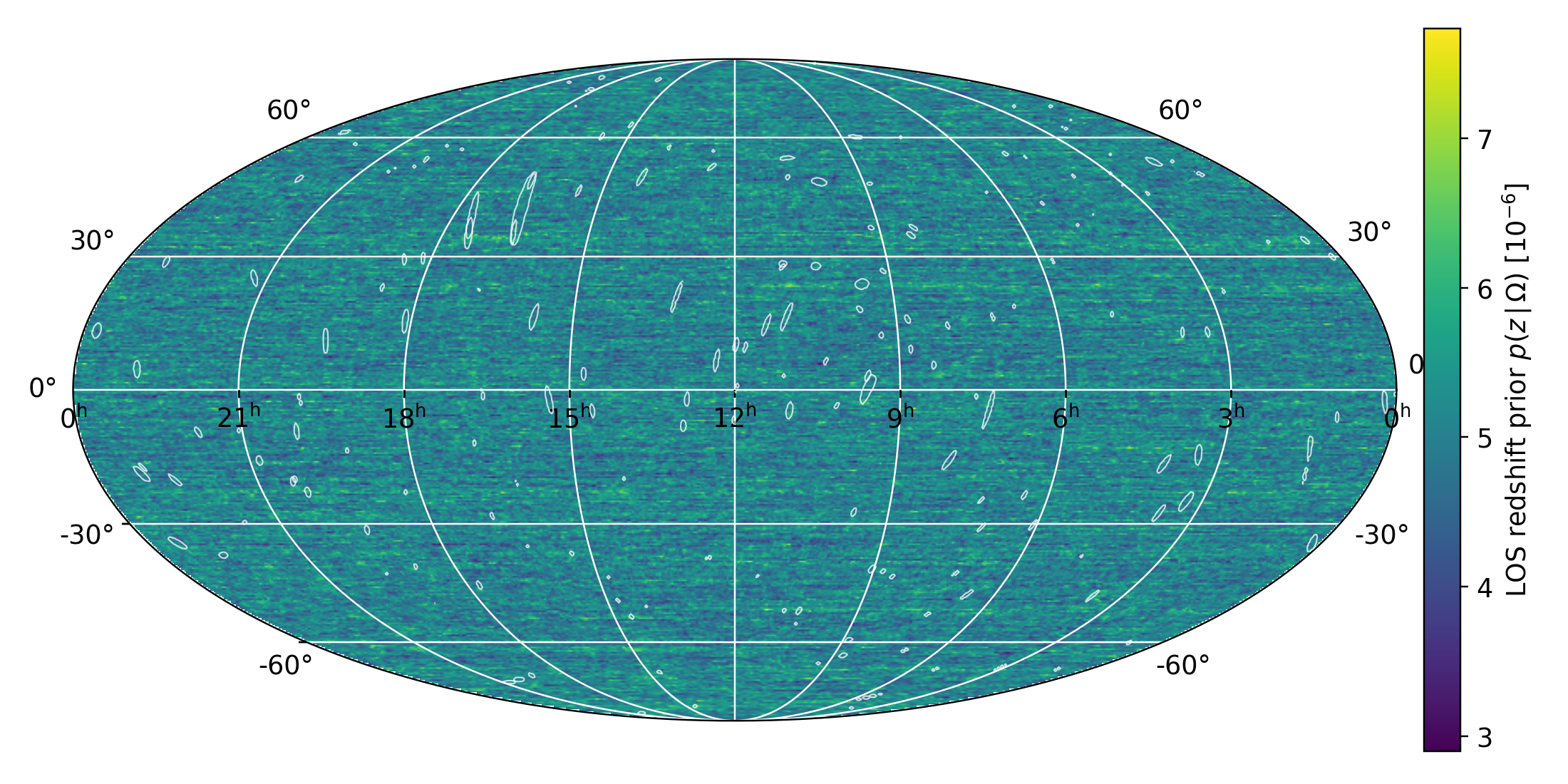}
    \caption{Sky localizations (90\% credible regions, represented by white contours) of the GW events used in the cosmological analysis, shown in equatorial coordinates (International Celestial Reference System). The background colour map shows the LOS redshift prior, marginalized over individual HEALPix pixels, computed in the $B$-band while accounting for magnitude uncertainties. The localization areas reflect the low-redshift mock-event setup and the controlled simulation conditions used in this work, and should not be interpreted as representative of the full population of real O4-like detections.}
    \label{fig:skyloc}
\end{figure*}

Fig.$~$\ref{fig:skyloc} shows the sky localizations (90\% credible regions) of the GW events used in our cosmological analysis, projected in equatorial coordinates. The localization areas are smaller than those of typical real GW detections, which mainly reflects the idealized mock-event setup used in this work: the events are drawn from a low-redshift mock galaxy catalogue and are analysed under controlled simulation conditions. The resulting sample should therefore not be interpreted as representative of the full population of O4-like detections, but as an idealized set designed to isolate the impact of catalogue-related uncertainties. In such cases, the information from the galaxy catalogue plays a dominant role in shaping the cosmological posterior. Consequently, we expect the impact of catalogue-related systematics—such as magnitude uncertainties or K-corrections—to be more significant than in poorly localized events, where the posterior averages over a larger number of potential host galaxies.

Using the modified \texttt{gwcosmo} package, we created several LOS redshift priors with various configurations of redshift and magnitude uncertainties, ranging from idealized to more realistic scenarios. We performed the tests using both $B$-band and $K$-band magnitudes, which are commonly used as luminosity-based proxies related to star formation and stellar mass, respectively. As a baseline, we created a reference run where redshift uncertainties were set to an extremely low fixed value of $10^{-6}$, simulating nearly perfect spectroscopic measurements. To model realistic spectroscopic uncertainties, we assigned redshift errors drawn from a Gaussian distribution with a mean of \mbox{$10^{-4} \times (1 + z)$} and a standard deviation of 10\% of this value, mimicking the scatter expected in high-quality spectroscopic datasets. To ensure positivity while preserving the shape of the original Gaussian, we sampled from a truncated normal distribution. In this model, the factor $10^{-4}$ sets the typical scale of the redshift error, which corresponds to a velocity uncertainty of $\mathord{\sim}30$ km/s in the local Universe ($z\lesssim0.01$). This choice is consistent with the statistical uncertainties reported by the SDSS spectroscopic pipeline, which are typically a few tens of km/s for galaxies \citep{SDSS_speczunc}, and is of the same order as the spectroscopic errors assumed by \cite{cross-parkin2025}.

\begin{figure}
    \centering
    \includegraphics[width=1\linewidth]{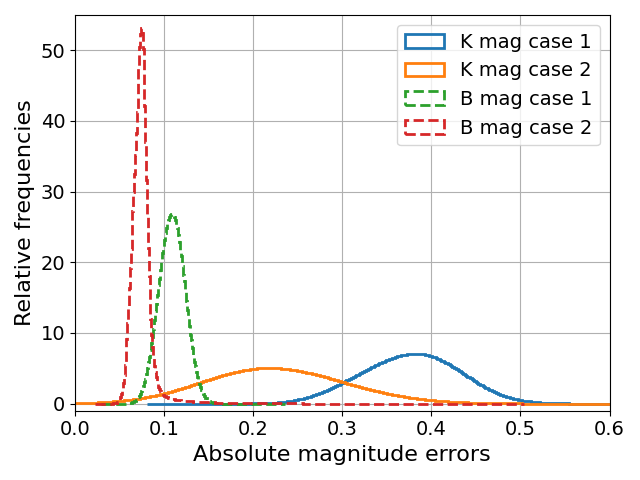}
    \caption{Histograms of absolute magnitude uncertainties adopted in the two prescriptions for the $B$- and $K$-bands. For case 1, the errors were modelled as truncated normal distributions with means set to rounded estimates of the mean plus two standard deviations of the GLADE+ error distributions and widths equal to 10\% of the mean. For case 2, the uncertainties were directly sampled from the error distributions of the GLADE+ catalogue (KDE for $B$, Gaussian fit for $K$). These models were used to propagate magnitude errors into the LOS redshift prior construction.}
    \label{fig:magerrs}
\end{figure}

To study the impact of magnitude uncertainties on the $H_0$ inference, we extended the spectroscopic-redshift case by including photometric magnitude uncertainties in both the $B$- and $K$-bands. We considered two prescriptions for assigning these uncertainties to the mock galaxies, both based on the observed magnitude errors in the GLADE+ catalogue \citep{GLADE+}. In the first approach (magnitude uncertainty case 1), we modelled the uncertainties as truncated normal distributions, with the mean set to 0.5\% of the magnitude in the $B$-band and 2\% in the $K$-band, and the standard deviation equal to 10\% of the corresponding mean, chosen to roughly represent the typical fractional variation of magnitude uncertainties seen in the GLADE+ catalogue, while slightly overestimating them to conservatively test their impact. These mean values were chosen as rounded estimates of the mean plus two standard deviations of the GLADE+ error distributions. The distributions were truncated between $10^{-6}$ and 1 to ensure physically meaningful values, and the resulting uncertainties were propagated into the LOS redshift prior construction via Eq.$~$(\ref{eq:LOS}). In the second, more realistic setup (magnitude uncertainty case 2), the uncertainties were directly sampled from the GLADE+ error distributions: for the $B$-band we reconstructed the empirical distribution using kernel density estimation, while for the $K$-band we fitted a Gaussian with mean $\mu_K = 0.012\, m_K$ and standard deviation $\sigma_K = 0.004\, m_K$, where $m_K$ is the apparent magnitude in the $K$-band. The resulting error distributions for the two prescriptions in both bands are shown in Fig.~\ref{fig:magerrs}.

To investigate the impact of K-corrections on the $H_0$ posterior, we performed tests in the $K$-band while neglecting magnitude uncertainties. In the $B$-band, the K-correction is typically low, and thus negligible for our mock catalogue. In contrast, in the near-infrared $K$-band, where the galaxy spectral slope is steeper and luminosity weighting is applied, even small redshift-dependent shifts can noticeably affect the inferred brightness. In the reference scenario with spectroscopic redshift errors,  we computed the apparent magnitudes from the absolute rest-frame magnitudes and the luminosity distance, according to
\begin{equation}
    m = M + 5\log_{10}\left(\frac{d_L}{10\,\text{pc}}\right),
\end{equation}
simulating the ideal case where K-corrections are perfectly applied. To test the effect of omitting K-corrections, we used the uncorrected apparent magnitudes from the catalogue, illustrating the resulting shift in the inferred posterior. Finally, we repeated the analysis with the built-in \texttt{gwcosmo} implementation, which applies a simplified correction in the $K$-band, \mbox{$K(z) = -6 \log_{10}(1+z)$} \citep{Kcorr}. In this case, we again used the catalogue magnitudes, but applied this correction in the analysis. Comparing these three configurations — ideal, no correction, and simplified empirical correction — enables us to quantify the systematic impact of neglecting or approximating K-corrections on the $H_0$ posterior. 

Throughout our tests we treated the mock catalogue as complete and therefore used only the in-catalogue contribution of Eq.$~$(\ref{eq:catparts}). In practice, real flux-limited catalogues require inclusion of the out-of-catalogue term, which further reduces the relative impact of the uncertainties studied here, meaning that our results represent an upper estimate of their expected effect. To minimize band-dependent differences introduced by the absolute-magnitude limits in \texttt{gwcosmo}, we adopted $M_{\max,K}=-15.79~\mathrm{mag}$ instead of the default value of $-19~\mathrm{mag}$, such that approximately $99.84\%$ of the mock galaxies are retained in both the $B$- and $K$-band analyses.

\section{Results}\label{sec:results}

\begin{figure}[htbp]
    \centering
    \includegraphics[width=1\linewidth]{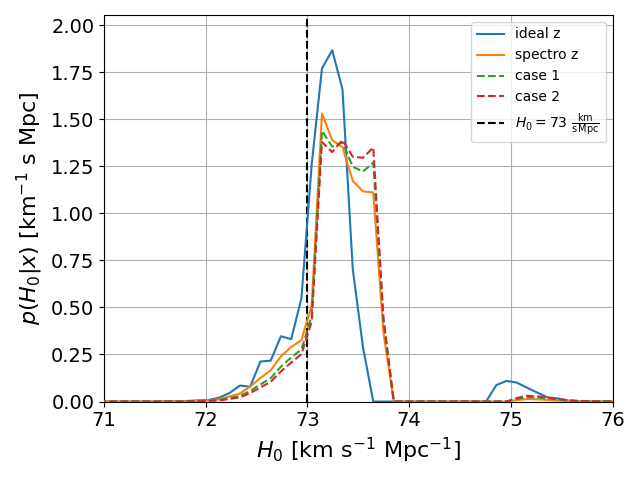}
    \includegraphics[width=1\linewidth]{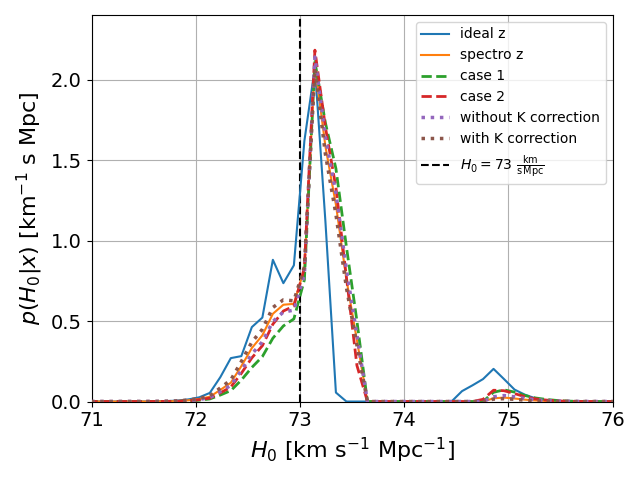}
    \caption{Posterior distributions of the Hubble constant obtained using the mock catalogue with different uncertainty treatments in the $B$ (top) and $K$ (bottom) bands. In the $B$-band we show results for the ideal and spectroscopic redshift cases, and the two prescriptions for magnitude uncertainties (cases 1 and 2; for details see Section~\ref{sec:methods}). In the $K$-band we include the reference scenarios with ideal and spectroscopic redshifts, the two magnitude uncertainty cases, and the runs with and without K-corrections. The vertical dashed line marks the reference value $H_0=73~\mathrm{km~s^{-1}~Mpc^{-1}}$, which was adopted during the generation of the mock galaxy catalogue. Since the resulting posteriors are visually very similar, we also present difference plots with respect to the reference runs (see Fig.$~$\ref{fig:diffs}).}
    \label{fig:overall}
\end{figure}

\begin{figure}[htbp]
    \centering
    \includegraphics[width=.96\linewidth]{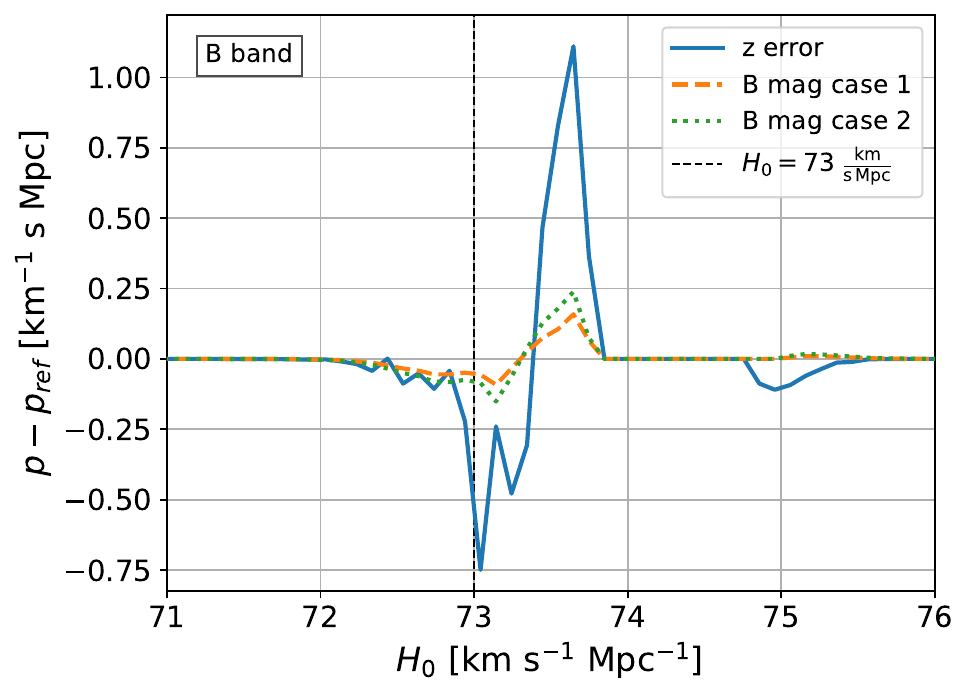}
    \includegraphics[width=.96\linewidth]{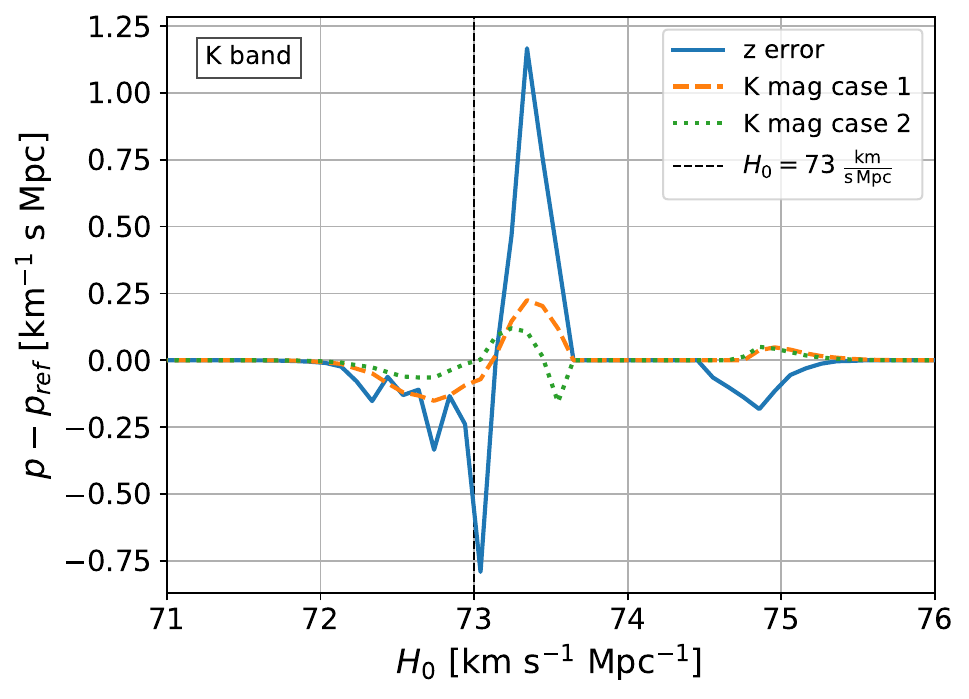}
    \includegraphics[width=.96\linewidth]{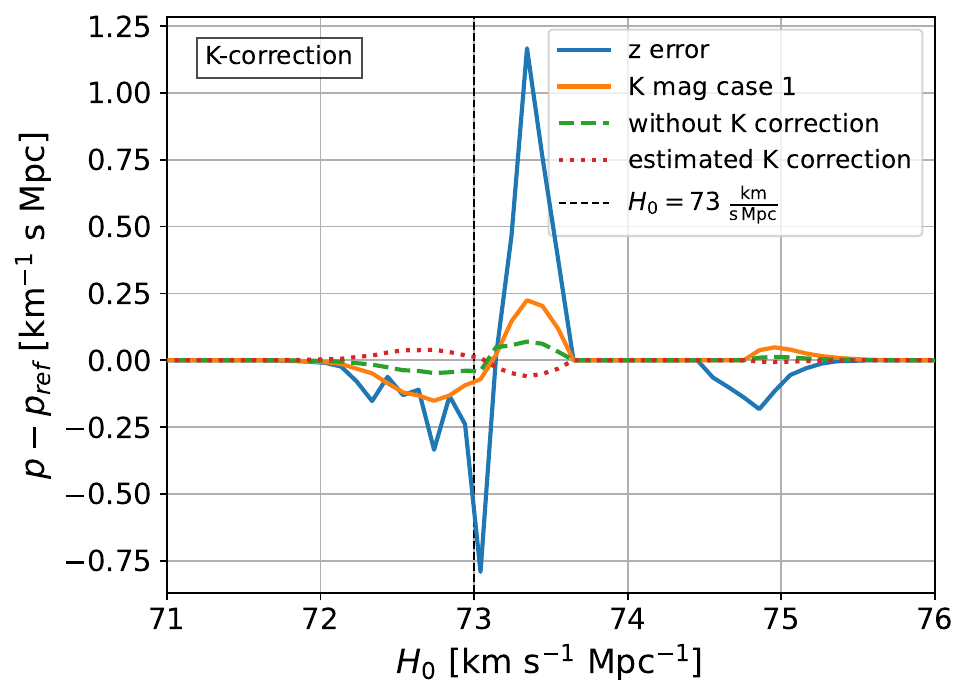}
    \caption{Differences of the Hubble constant posteriors with respect to the local reference baseline relevant for each tested effect. Local baselines, rather than a single global baseline, are used to isolate the incremental contributions of redshift errors, magnitude uncertainties, and K-correction assumptions. \text{Top:} $B$-band results, showing the impact of spectroscopic redshift errors relative to the ideal case and the impact of the two magnitude uncertainty prescriptions relative to the spectroscopic redshift baseline. \text{Middle:} $K$-band results, showing the impact of spectroscopic redshift errors relative to the ideal case and the impact of the two magnitude uncertainty prescriptions relative to the spectroscopic redshift baseline. \text{Bottom:} $K$-band results illustrating the incremental effects of spectroscopic redshift errors, magnitude uncertainty case$~$1, and omitting or estimating the K-corrections, using the corresponding $K$-band reference configurations as local baselines. In each panel the vertical dashed line marks the fiducial value $H_0 = 73~\mathrm{km\,s^{-1}\,Mpc^{-1}}$.}
    \label{fig:diffs}
\end{figure}

\begin{figure}
    \centering
    \includegraphics[width=1\linewidth]{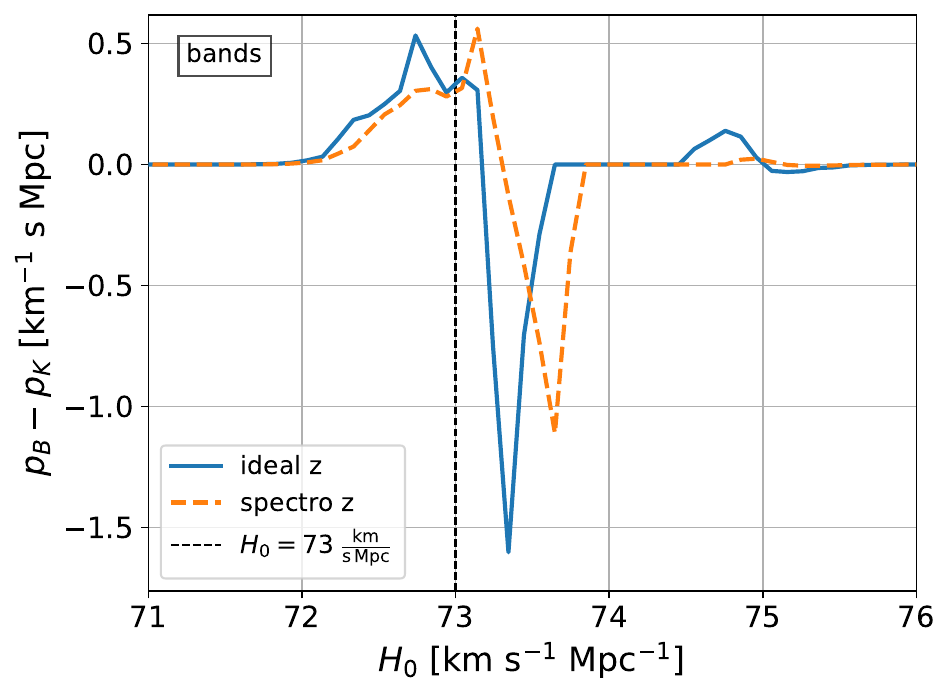}
    \caption{Differences between the $B$- and $K$-band $H_0$ posteriors, shown as $p_B(H_0)-p_K(H_0)$, for the ideal redshift and spectroscopic redshift cases.  The comparison illustrates the impact of the luminosity weighting scheme itself, separately from the uncertainty treatments shown in Fig.~\ref{fig:diffs}. The vertical dashed line marks the fiducial value $H_0 = 73~\mathrm{km\,s^{-1}\,Mpc^{-1}}$.}
    \label{fig:banddiff}
\end{figure}

\begin{figure}
    \centering
    \includegraphics[width=1\linewidth]{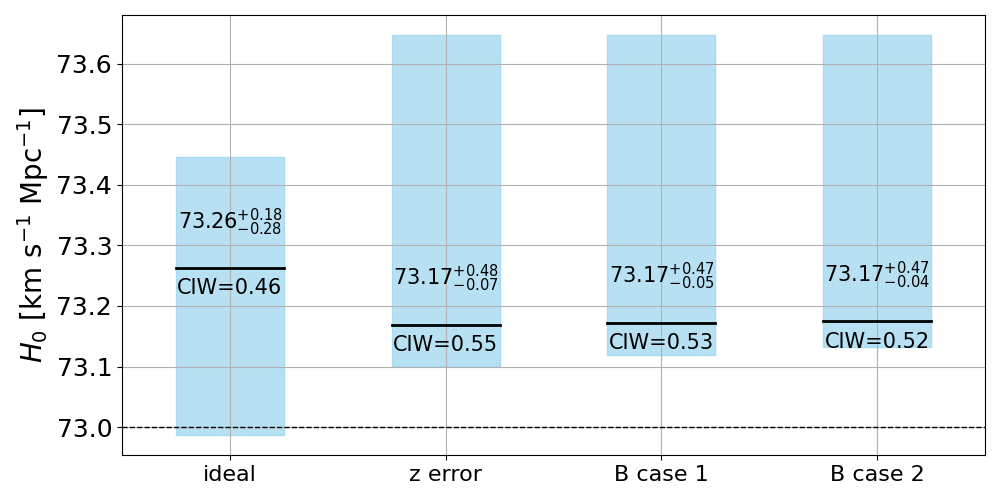}
    \includegraphics[width=1\linewidth]{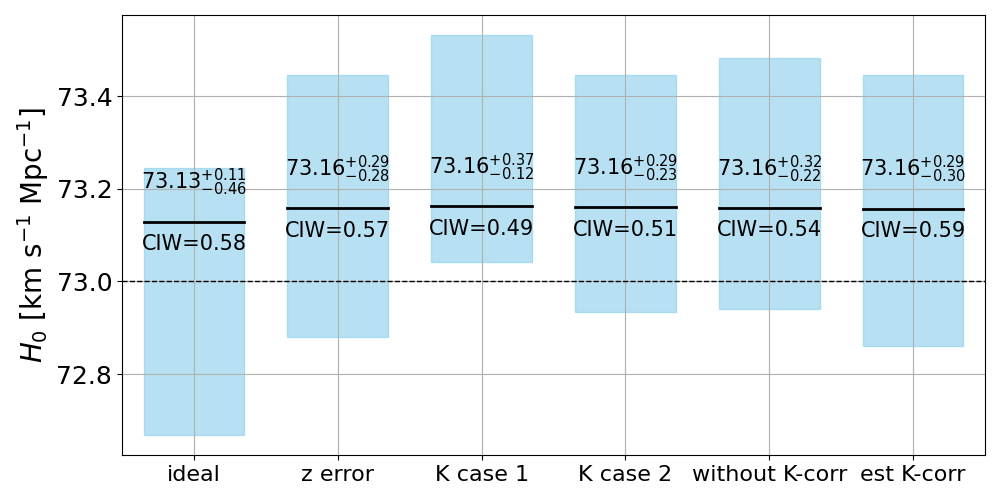}
    \caption{MAP estimates and 68\% HDIs for the Hubble constant obtained with different uncertainty treatments in the $B$ (top) and $K$ (bottom) bands. Blue boxes mark the 68\% HDIs, black bars indicate the MAP values estimated from spline-interpolated one-dimensional posteriors, and annotations report the numerical MAP values with the HDI edges as errors  together with the credible interval width (CIW).}
    \label{fig:HDI}
\end{figure}

We focus primarily on the impact of photometric magnitude uncertainties and K-correction assumptions on the inferred value of the Hubble constant, while including realistic spectroscopic redshift errors for comparison, to place these effects in context with a well-known source of uncertainty. We first show the overall posteriors (Fig.$~$\ref{fig:overall}), followed by their differences with respect to the corresponding reference runs (Fig.$~$\ref{fig:diffs}) and a direct comparison of the $B$- and $K$-band results (Fig.$~$\ref{fig:banddiff}). Finally, we summarize the maximum a posteriori (MAP) values and credible intervals (Fig.~\ref{fig:HDI}). Since our aim is to quantify the incremental impact of each uncertainty treatment, we focus on relative changes with respect to the corresponding reference runs rather than on the absolute displacement of any individual posterior from the input value.

Fig.$~$\ref{fig:overall} shows an overview of the $H_0$ posteriors obtained with the $B$- (top) and $K$- (bottom) band analyses. The vertical dashed line in both panels marks the reference value $H_0~=~ 73~\mathrm{km\,s^{-1}\,Mpc^{-1}}$, consistent with the cosmology adopted in the Millennium Simulation used to generate the mock galaxy catalogue. The finite mock-event sample, the discrete galaxy-catalogue realization, and the finite resolution of the grid-based likelihood evaluation lead to small-scale irregularities in the posterior shapes. Nevertheless, the posteriors remain concentrated close to the fiducial value in all tested configurations. In the $B$-band, the inclusion of spectroscopic redshift errors visibly broadens and reshapes the main posterior peak. In the $K$-band, the dominant posterior peak remains nearly fixed when spectroscopic redshift errors are included. The magnitude-uncertainty and K-correction treatments produce only minor changes in the posterior shapes.

Fig.$~$\ref{fig:overall} displays results with the $B$- and $K$-band luminosity weighting on the same $H_0$ axis range, allowing for a direct visual comparison. In our simulations, host galaxies were selected with probabilities proportional to both stellar mass and the merger-rate evolution with redshift \citep[following][]{MandD}. Since the mock catalogue extends only to $z < 0.2$, the evolution term is nearly constant, and stellar mass effectively dominates the host selection, defining the underlying distribution in our setup, as described in detail in \citet{UtilizingStellarMass}. Accordingly, $K$-band luminosity provides a good proxy for this weighting, while $B$-band luminosity emphasizes star formation, corresponding to a mismatched weighting scheme. This provides a natural interpretation of the band-dependent behaviour seen in Fig.~\ref{fig:overall}, as the dominant $K$-band posterior peak remains closer to the fiducial value than in the  $B$-band analysis.

Since most differences are subtle in Fig.$~$\ref{fig:overall}, we also present the incremental posterior differences in Fig.$~$\ref{fig:diffs}. Rather than using a single global baseline for all curves, each comparison is made with respect to the corresponding reference run that excludes the uncertainty being tested. This choice allows us to separate the impact of spectroscopic redshift errors from the additional effects of magnitude uncertainties and K-correction assumptions. In the $B$- and $K$-band panels, the spectroscopic-redshift case is compared to the ideal-redshift case, while the magnitude-uncertainty cases are compared to the corresponding spectroscopic-redshift baseline. Thus, the magnitude curves show the additional effect of magnitude uncertainties after realistic spectroscopic redshift errors have already been included. In the K-correction panel, the comparisons are defined analogously with respect to the $K$-band reference configuration. 

In the top panel of Fig.~\ref{fig:diffs}, consistent with the overall posterior behaviour seen in Fig.~\ref{fig:overall}, spectroscopic redshift errors lead to noticeably larger deviations from the ideal case than those induced by magnitude uncertainties, which produce only minor changes relative to the spectroscopic baseline. A similar trend is seen in the $K$-band (middle panel). When comparing the two magnitude-uncertainty prescriptions, their relative effect is reversed between the two bands: in the $K$-band case 1 produces slightly larger deviations than case 2, while in the $B$-band the opposite trend is observed. As illustrated in Fig.$~$\ref{fig:magerrs}, this counter-intuitive behaviour can be understood from the fact that in case$~$2 most galaxies have smaller errors, but a small fraction ($\leq 1\%$) exhibit much larger uncertainties, and their effect can strengthen in the LOS redshift prior. The bottom panel of Fig.$~$\ref{fig:diffs} focuses on the treatment of K-corrections in the $K$-band, with the spectroscopic redshift errors and the $K$-band magnitude uncertainty case 1 included for comparison. The omission of K-corrections produces a smaller deviation from the corresponding reference posterior than either the inclusion of spectroscopic redshift errors or the tested magnitude uncertainties. Applying the empirical K-correction changes the sign of the residual difference over part of the posterior, indicating a different redistribution of probability relative to the reference case.

To quantify the deviations shown visually in Figs.~\ref{fig:overall} and \ref{fig:diffs}, we computed the first Wasserstein distance (also known as the Earth Mover’s Distance) between each posterior and its corresponding reference run (Table~\ref{tab:Wasserstein}). This metric quantifies the minimal “effort” required to transform one probability distribution $p_1(H_0)$ into another $p_2(H_0)$ by integrating their cumulative differences:
\begin{equation}
    W_1(p_1, p_2) = \int \mid F_1(H_0)-F_2(H_0)\mid \mathrm{d} H_0,
\end{equation}
where $F_1$ and $F_2$ denote the corresponding cumulative distribution functions. Larger $W_1$ values indicate greater overall differences between the posteriors in both location and shape (see, e.g. \citet{vaserstein, rubner, Panaretos}). The difference curves in Fig.~\ref{fig:diffs} provide a visual counterpart to this comparison: positive and negative residuals indicate how probability is redistributed across the $H_0$ range. The Wasserstein distance then quantifies the corresponding integrated redistribution through the cumulative distributions, so that deviations of opposite sign do not simply cancel.

The Wasserstein distances quantify the trends identified in the posterior and difference plots. Spectroscopic redshift errors produce the largest integrated posterior differences in both bands, while the effects of magnitude uncertainties are smaller and band dependent. The reversed ordering of the two magnitude-uncertainty prescriptions between the $B$- and $K$-bands is also reflected in the Wasserstein distances, supporting the interpretation that their impact depends not only on the typical error size, but also on the shape of the error distribution and its interplay with the luminosity weighting. The K-correction treatments produce the smallest integrated differences, with the empirical correction yielding a slightly smaller $W_1$ than omitting the correction.

We next assess the effect of the luminosity-weighting scheme by directly comparing the $B$- and $K$-band posteriors in Fig.~\ref{fig:banddiff}. The plotted differences, $p_B(H_0)-p_K(H_0)$, show that the choice of band changes the posterior shape already in the ideal redshift case, with a Wasserstein distance of $W_1 = 0.224~\mathrm{km,s^{-1},Mpc^{-1}}$. After including spectroscopic redshift errors, the corresponding distance remains similar, $W_1 = 0.204~\mathrm{km,s^{-1},Mpc^{-1}}$. These values are comparable to the largest integrated changes produced by the tested uncertainty treatments, showing that the luminosity-weighting scheme is an important modelling choice in catalogue-based dark-siren inference. This sensitivity to the luminosity-weighting prescription is consistent with previous studies \citep[e.g.][]{Perna, SM_SFR, Alfradique_2025}, which showed that mismatched host-galaxy weighting models can affect the location and width of the inferred $H_0$ posterior.

Because the Wasserstein distance integrates cumulative differences over the full posterior, it does not necessarily emphasize local peak-region changes in the same way as a visual comparison of the posterior curves. We therefore also examine the MAP values and credible intervals to assess how the tested uncertainty treatments affect the peak locations and posterior widths.

\begin{table}[htbp]
      \caption[]{Wasserstein distances between the $H_0$ posteriors and the corresponding reference runs, for different uncertainty treatments in the $B$- and $K$-bands. The distances are given in $\mathrm{km~s^{-1}~Mpc^{-1}}$. The reference is the ideal case for the redshift errors, and the spectroscopic-redshift baseline for the magnitude-uncertainty and K-correction cases, consistent with the comparisons shown in Fig.$~$\ref{fig:diffs}.}
         \label{tab:Wasserstein}
     $$ 
         \begin{array}{p{0.5\linewidth}c c}
            \hline
           \noalign{\smallskip}
                  &  B & K \\
            \noalign{\smallskip}
            \hline
            \noalign{\smallskip}
            z error  & 0.178 & 0.210    \\
            case 1 & 0.042 & 0.096           \\
            case 2 & 0.065 & 0.050  \\
            without K-correction & - & 0.029           \\
            estimated K-correction & - & 0.022          \\
            \noalign{\smallskip}
            \hline
        \end{array}
     $$ 
\end{table}

Fig.~\ref{fig:HDI} summarizes the MAP estimates and 68\% highest-density credible intervals (HDIs) for all uncertainty models. In the $B$-band, the inclusion of spectroscopic redshift errors shifts the MAP slightly towards lower $H_0$, while the 68\% HDI extends further towards higher $H_0$, consistent with the broadening and flattening of the main posterior peak seen in Fig.~\ref{fig:overall}. The magnitude-uncertainty cases leave both the MAP and the interval widths nearly unchanged relative to the spectroscopic-redshift baseline. In the $K$-band, the stability of the dominant posterior peak seen in Fig.~\ref{fig:overall} is reflected in the nearly unchanged MAP values, while the uncertainty treatments mainly affect the HDI boundaries and widths. Among the magnitude-uncertainty cases, case 1 produces the most noticeable change. These changes are not monotonic: adding magnitude uncertainties or changing the K-correction treatment can slightly narrow or broaden the HDI, reflecting changes in the detailed posterior shape rather than a simple increase in the overall uncertainty.

Overall, our results show that spectroscopic redshift errors cause the dominant effect in our setup, while realistic magnitude uncertainties and K-correction treatments have only a minor impact. The direct comparison of the $B$- and $K$-band results indicates that the choice of luminosity-weighting scheme can affect the $H_0$ posterior at a level comparable to the largest uncertainty-induced changes.

\FloatBarrier
\section{Conclusions}\label{sec:conclusions}

Here we list our main findings regarding the impact of the studied catalogue-related uncertainties and luminosity-weighting choices on the inference of the Hubble constant from dark sirens. In all tested configurations, the resulting $H_0$ posteriors remain close to the fiducial value used in the mock-catalogue generation, and the effects studied here are small in absolute terms.

\begin{enumerate}

    \item Weighting scheme. The choice of luminosity-weighting scheme has a substantial impact on the $H_0$ posterior. The direct comparison between the $B$- and $K$-band analyses shows differences that are comparable to the largest integrated changes induced by the tested uncertainty treatments. The dominant peak of the $K$-band posterior remains closer to the fiducial $H_0$ value than in the $B$-band analysis, as $K$-band luminosity traces stellar mass more closely and thus better matches the host-galaxy weighting used in our mock setup.
    
    \item Redshift uncertainties. Among the tested uncertainty treatments, spectroscopic redshift errors produce the largest integrated changes in the $H_0$ posterior in both the $B$- and $K$-band analyses. In the $K$-band this integrated change is large even though the dominant posterior peak remains visually more stable than in the $B$-band, showing that peak shifts alone do not fully characterize the impact of redshift-error modelling. Integrated measures such as the Wasserstein distance therefore provide a more complete characterization of changes in the full posterior shape.

    \item Magnitude uncertainties.     Photometric magnitude uncertainties have a smaller effect than the redshift errors, and their impact is band dependent.  In the $B$-band, magnitude-uncertainty case 2 leads to a larger Wasserstein distance, whereas in the $K$-band case 1 gives the larger effect.     This shows that the impact of magnitude errors is not determined solely by their typical size, but also depends on the shape of the error distribution and on its interplay with the adopted luminosity weighting.

    \item K-corrections.  Omitting K-corrections in the $K$-band produces a smaller integrated deviation from the reference posterior than both spectroscopic redshift errors and the tested magnitude-uncertainty prescriptions. Applying the empirical K-correction slightly reduces this integrated difference relative to the no-correction case, although small residual differences remain in the posterior shape. 
    
\end{enumerate}
   
In this work we neglected the out-of-catalogue contribution and assumed a complete, volume-limited galaxy sample up to $z<0.2$. While real analyses typically involve larger localization areas, our simulated events are relatively well localized and therefore represent a favourable limiting case of sky localization. This approach allows us to isolate the effects of galaxy-catalogue systematics without the additional complexity introduced by broad sky maps. Real galaxy catalogues are flux-limited, and photometric or mixed redshift errors will often dominate over spectroscopic ones. Under such conditions, the relative impact of magnitude uncertainties is expected to diminish even further. We therefore conclude that magnitude errors need not be a priority in dark siren cosmology in the near future, whereas the accurate treatment of redshift errors remains essential. Under the conditions considered here, K-correction assumptions likewise have only a minor impact. At the same time, our results highlight the importance of adopting a physically motivated host-galaxy weighting scheme in catalogue-based dark-siren analyses.

\begin{acknowledgements}
This work makes use of gwcosmo which is available at \url{https://git.ligo.org/lscsoft/gwcosmo}.

Data used in this work was generated using Swinburne University's Theoretical Astrophysical Observatory (TAO). TAO is
part of the Australian All-Sky Virtual Observatory (ASVO) and is freely accessible at \url{https://tao.asvo.org.au/tao/}.
The Millennium Simulation was carried out by the Virgo Supercomputing Consortium at the Computing Centre of the Max
Planck Society in Garching. It is publicly available at \url{http://www.mpa-garching.mpg.de/Millennium/}.
The Semi-Analytic Galaxy Evolution (SAGE) model used in this work is a publicly available codebase that runs on the dark
matter halo trees of a cosmological N-body simulation. It is available for download at
\url{https://github.com/darrencroton/sage}.

The authors are grateful for computational resources provided by the LIGO Laboratory and supported by  National Science Foundation's grant Nos. PHY-0757058 and PHY-0823459.

This project has received funding from the HUN-REN Hungarian Research Network and was also supported by the NKFIH excellence grant TKP2021-NKTA-64.

The authors acknowledge the use of OpenAI's ChatGPT for assistance in refining the language and improving the clarity of the manuscript.
\end{acknowledgements}

%
%________________________________________________________________

\section*{Data availability}
The full analysis pipeline developed for this work is publicly available at
\url{https://github.com/MariaPalfi/dark-siren-magnitude-systematics}.
The simulation outputs and large data products required to reproduce the results are archived at Zenodo:
\url{https://doi.org/10.5281/zenodo.15054067}.
Additional data supporting the findings of this study are available from the corresponding author upon reasonable request.

\bibliographystyle{aa}
\bibliography{references}

\end{document}